\def\pr#1 {Phys. Rev. {\bf D#1\tie \rm }}
\def\pe#1 {Phys. Rev. {\bf #1\tie\rm }}
\def\pre#1 {Phys. Rep. {\bf #1\tie\rm }}
\def\pl#1 {Phys. Lett. {\bf #1B\tie \rm }}
\def\prl#1 {Phys. Rev. Lett. {\bf #1\tie \rm }}
\def\np#1 {Nucl. Phys. {\bf B#1\tie \rm }}
\def\ap#1 {Ann. Phys. (NY) {\bf #1\tie \rm }}
\def\cmp#1 {Commun. Math. Phys. {\bf #1\tie \rm }}
\def\imp#1 {Int. Jour. Mod. Phys. {\bf A#1\tie \rm }}
\def\mpl#1 {Mod. Phys. Lett. {\bf A#1\tie\rm }}
\def\jhep#1 {JHEP {\bf #1\tie\rm }}
\def\zp#1 {Z. Phys. {\bf C#1\tie\rm }}

\def\tie{\noexpand~}

\def\be{\begin{equation}}
\def\ee{\end{equation}}
\def\bea{\begin{eqnarray}}
\def\eea{\end{eqnarray}}

\catcode`@=11
\def\marginnote#1{}
\newcount\hour
\newcount\minute
\newtoks\amorpm
\hour=\time\divide\hour by60
\minute=\time{\multiply\hour by60 \global\advance\minute by-\hour}
\edef\standardtime{{\ifnum\hour<12 \global\amorpm={am}%
        \else\global\amorpm={pm}\advance\hour by-12 \fi
        \ifnum\hour=0 \hour=12 \fi
        \number\hour:\ifnum\minute<10 0\fi\number\minute\the\amorpm}}
\edef\militarytime{\number\hour:\ifnum\minute<10 0\fi\number\minute}
\def\draftlabel#1{{\@bsphack\if@filesw {\let\thepage\relax
   \xdef\@gtempa{\write\@auxout{\string
      \newlabel{#1}{{\@currentlabel}{\thepage}}}}}\@gtempa
   \if@nobreak \ifvmode\nobreak\fi\fi\fi\@esphack}
        \gdef\@eqnlabel{#1}}
\def\@eqnlabel{}
\def\@vacuum{}
\def\draftmarginnote#1{\marginpar{\raggedright\scriptsize\tt#1}}
\def\draft{\oddsidemargin 0.0truein
        \def\@oddfoot{\sl preliminary draft \hfil
        \rm\thepage\hfil\sl\today\quad\militarytime}
        \let\@evenfoot\@oddfoot \overfullrule 3pt
        \let\label=\draftlabel
        \let\marginnote=\draftmarginnote
   \def\@eqnnum{(\theequation)\rlap{\kern\marginparsep\tt\@eqnlabel}%
\global\let\@eqnlabel\@vacuum}  }
\catcode`@=12

\documentstyle[epsf, 12pt]{article}
\begin{document}

\thispagestyle{empty}
\begin{flushright}
RU05-03-B
\\
hep-th/0507306n
\end{flushright}

\bigskip\bigskip
\begin{center}
\Large{\bf A NEUTRAL TWO-FLAVOR LOFF 
COLOR SUPERCONDUCTOR}
\end{center}

\vskip 1.0truecm

\centerline{\bf Ioannis
Giannakis${}^{a}$, De-fu Hou${}^{b}$ and Hai-Cang Ren${}^{a, b}$
\footnote{giannak@summit.rockefeller.edu,
hdf@iopp.ccnu.edu.cn,
ren@summit.rockefeller.edu}}
\vskip5mm

\centerline{${(a)}${\it Physics Department, Rockefeller University, }}
\centerline{\it New York, NY 10021, U. S. A.}
\centerline{${(b)}${\it Institute of Particle Physics,
Central China Normal University,}}
\centerline{\it Wuhan, 430079, China}

\vskip5mm

\bigskip \nopagebreak \begin{abstract}
\noindent
In this paper we construct analytically a LOFF color
superconducting state that is both
color and charge neutral using the weak coupling approximation.
We demonstrate that this
state is free from chromomagnetic instabilities. Its relevance 
to the realistic quark matter at moderately high baryon density 
is discussed.

\end{abstract}

\newpage\setcounter{page}1

\vfill\vfill\break

The ground state of quark matter at moderately high baryon
density and very low temperature
has been an active area of research recently because of its
relevance to the interior
of a compact star. While a color superconductor \cite{BL} is
expected, it
is unlikely to be of the conventional BCS type
because the pairing occurs
between  quarks of different flavors and Fermi momenta.
The physics of Cooper pairing among unequal Fermi momenta
applies also in systems of cold atomic gasses \cite{LW} as well. 
Several exotic color superconducting states 
have been proposed as candidate ground states. These
include the gapless CSC state \cite{SH}\cite{HZC}\cite{AKR}, 
the LOFF-CSC state \cite{ABR}, and a state that consists of a 
heterogeneous mixture of gapped CSC and normal phases \cite{BCR}.
But none is completely satisfactory for different reasons. The gapless CSC 
state-a generalization of the Sarma state in solid state physics-although
being stable against small variations of the gap parameters and with the
condition
of local charge neutrality fully implemented, it suffers from chromomagnetic 
instabilities when coupled to the gluon-photon field \cite{HS}
\cite{CGMNR}\cite{GR1}\cite{AW}\cite{GR2}\cite{H}. In contrast,
the LOFF state, although
is free from chromagnetic instabilities \cite{GR2} the charge 
neutrality condition 
has to be implemented and its free energy may favor a complicated 
superposition of plane waves, which makes even numerical treatment 
difficult \cite{BR}. The mixed phase scenario, being an alternative 
way to avoid the problem of chromomagnetic instabilities, 
has failed to make quantitative 
comparisons between its free energy with its homogeneous competitors.    

In a previous paper\cite{GR2}, we demonstrated the chromomagnetic stability of 
a simple two-flavor LOFF state
consisting of a single plane wave, within the region of the displacement 
parameter where it is energetically favored compared to the
BCS and normal phases.
The analysis was carried out using a grand canonical ensemble at fixed 
baryon and electric chemical potentials while the 8th-color 
chemical potential was set to zero. In this paper, we shall 
construct analytically a  
two flavor LOFF state using the canonical ensemble at fixed baryon density and 
we shall implement
the color and charge neutrality conditions. In order to justify 
the weak coupling approximation that we shall employ, 
we shall treat the sum of the electric 
charges of the u and d quarks
as free parameter $\epsilon$, more specifically we write
\begin{equation}
Q_u=\frac{1}{3}+\epsilon \qquad Q_d=-\frac{1}{3}.
\label{first}
\end{equation}
The accuracy of the weak coupling approximation requires that $\epsilon<<1$, 
but we expect that the qualitative conclusions as to the existence of a
neutral and stable LOFF state drawn in this paper can be 
extrapolated to the realistic case $\epsilon=\frac{1}{3}$ of the real world. 

The dynamics of quark matter with two flavors can be described by the 
following NJL effective Lagrangian
\begin{eqnarray}
{\cal L} = -\bar\psi\gamma_\mu\frac{\partial\psi}{\partial x_\mu}
+\bar\psi\gamma_4\mu\psi
&+&G_S[(\bar\psi\psi)^2+(\bar\psi\vec\tau\psi)^2]\nonumber \\
&+&G_D(\bar\psi_C\gamma_5\epsilon^c\tau_2\psi)
(\bar\psi\gamma_5\epsilon^c\tau_2\psi_C),
\label{second}
\end{eqnarray}
where $\psi$ represents the quark fields and $\psi_C=C\tilde{\bar\psi}$
its charge conjugate with $C=i\gamma_2\gamma_4$. All gamma 
matrices are hermitian,
$(\epsilon^c)^{mn}=\epsilon^{cmn}$ is 
a $3\times 3$ matrix acting on the red(r), green(g) and blue(b) 
color indices and the Pauli matrices $\vec\tau$ act on the $u$ and $d$ 
flavor (isospin) indices.

The chemical potential written in a matrix form
in the color-flavor space reads
\begin{equation}
\mu = \frac{1}{3}\mu_B+\mu_QQ+\mu_8\lambda_8,
\label{third}
\end{equation}
or in component form
\begin{eqnarray}
\mu_u^{r,g} &=& \frac{1}{3}\mu_B+\Big(\frac{1}{3}
+\epsilon\Big)\mu_Q+\frac{1}{2\sqrt{3}}\mu_8,\\ \nonumber
\mu_d^{r,g} &=& \frac{1}{3}\mu_B-\frac{1}{3}\mu_Q
+\frac{1}{2\sqrt{3}}\mu_8,\\ \nonumber
\mu_u^{b} &=& \frac{1}{3}\mu_B+\Big(\frac{1}{3}+\epsilon\Big)\mu_Q
-\frac{1}{\sqrt{3}}\mu_8,\\ \nonumber
\mu_d^{b} &=& \frac{1}{3}\mu_B-\frac{1}{3}\mu_Q-\frac{1}{\sqrt{3}}\mu_8,
\label{forth}
\end{eqnarray}
where $\mu_B$ denotes the chemical potential associated
with the baryon number, $\mu_Q$ with the electric 
charge and $\mu_8$ with the eighth color. In terms of the notation of Ref. 
\cite{GR2}, i.e. $\mu = \bar\mu-\delta\tau_3+\delta^\prime\lambda_8$. 
We have
\begin{eqnarray}
\bar\mu &=& \frac{1}{3}\mu_B+\frac{1}{2}\epsilon\mu_Q,\\ \nonumber
\delta &=& -\Big(\frac{1}{3}+\frac{1}{2}\epsilon\Big)\mu_Q
\simeq -\frac{1}{3}\mu_Q,\\ \nonumber
\delta^\prime &=& \mu_8,
\label{fifth} 
\end{eqnarray}
where $\bar\mu=\frac{1}{6}{\rm tr}\mu$ denotes the mean chemical 
potential of all colors and flavors, $\delta$ measures 
the displacement between the Fermi momenta of the two flavors and 
$\delta^\prime$ is 
proportional to the displacement between the blue color and the other colors. 

In the presence of a diquark condensate, 
\begin{equation}
<\bar\psi(\vec r)\gamma_5\lambda_2\tau_2\psi_C(\vec r)>
=\Phi e^{2i\vec q\cdot\vec r},
\label{sixth}
\end{equation}
the pressure at $T=0$ can be
calculated using the mean field approximation. We find that 
\begin{equation}
p = p_n + p_{\rm cond.}
\label{seventh}
\end{equation}
where $p_n$ refers to the contribution of the
normal phase and is approximated by an ideal 
gas of quarks, i.e.
\begin{eqnarray}
& p_n & = \frac{1}{12\pi^2}{\rm tr}\mu^4=\nonumber \\
& \frac{1}{12\pi^2} & \Big[2\Big(\bar\mu-\delta
+\frac{\delta^\prime}{2\sqrt{3}}\Big)^4
+2\Big(\bar\mu+\delta+\frac{\delta^\prime}{2\sqrt{3}}\Big)^4
+\Big(\bar\mu-\delta-\frac{\delta^\prime}{\sqrt{3}}\Big)^4
+\Big(\bar\mu+\delta-\frac{\delta^\prime}{\sqrt{3}}\Big)^4\Big].
\label{eighth}
\end{eqnarray}
The contribution from the condensate, $p_{\rm cond.}$ was
computed in Ref. \cite{GR2} for $\delta^\prime=0$
in the weak coupling approximation
\begin{eqnarray}
p_{\rm cond.} &=&-\Gamma(\bar\mu,\Delta,\delta,q)\nonumber \\
&\simeq& -\frac{2{\bar\mu}^2}{\pi^2}\lbrace\Delta^2\Big
(\ln\frac{\Delta}{\Delta_0}
-\frac{1}{2}\Big)+\frac{(q+\delta)^3}{4q}
\Big[(1-x_1^2)\ln\frac{1+x_1}{1-x_1}
+\frac{2}{3}(2x_1^3-3x_1+1)\Big]\nonumber \\
&+&\frac{(q-\delta)^3}{4q}\Big[(1-x_2^2)
\ln\frac{1+x_2}{1-x_2}+\frac{2}{3}(2x_2^3-3x_2+1)\Big]\rbrace,
\label{nineth}
\end{eqnarray}
where $x_1$ and $x_2$ are dimensionless parameters that
were introduced in 
Ref. \cite{TI}, i.e.
\begin{equation}
x_1=\theta\Big(1-\frac{\Delta}{(q+\delta)}\Big)
\sqrt{1-\frac{\Delta^2}{(q+\delta)^2}}
\label{tenth}
\end{equation}
\begin{equation}
x_2=\theta\Big(1-\frac{\Delta}{|q-\delta|}\Big)
\sqrt{1-\frac{\Delta^2}{(q-\delta)^2}}.
\label{eleventh}
\end{equation}
and $\Delta_0$ is the BCS gap energy given by the expression 
\begin{equation}
\frac{1}{4G_D}=\frac{2\bar\mu^2}{\pi^2}\ln\frac{2\omega_0}{\Delta_0}
\label{twelveth}
\end{equation}
where $\omega_0$ represents
the UV cutoff of the pairing force. In a grand canonical
ensemble approach, the pressure ought to be maximized with respect to $\Delta$ 
and $q$ at equilibrium.
The weak coupling approximation is justified if
the conditions that $\Delta_0<<\mu_0$ and $\Delta$, $\delta$,
and $q$ are comparable or smaller than $\Delta_0$
are satisfied. Since we shall employ the weak coupling 
approximation, only terms up to the order of $\bar\mu^2\Delta^2$ are retained 
in the expression for the pressure.

The equilibrium conditions
\begin{equation}
\Big(\frac{\partial p}{\partial\Delta^2}\Big)_{\bar\mu,\delta,\delta^\prime,q}
=0, \qquad
\Big(\frac{\partial p}
{\partial q}\Big)_{\bar\mu,\delta,\delta^\prime,\Delta}=0,
\label{thirteenth}
\end{equation}
allow three types of CSC solutions:

1) The 2SC state ( BCS state ). This solution is characterized
by $q=0$ and $\Delta=\Delta_0$. The
contribution of the condensate to the pressure is given by 
\begin{equation}
p_{\rm cond.}=\frac{\bar\mu^2}{\pi^2}(\Delta_0^2-2\delta^2).
\label{fourteenth}
\end{equation}
This solution corresponds to the range  of the displacement 
parameter $\delta<\Delta_0$.

2) The g2SC state, which is
characterized by $q=0$, and $\Delta=\sqrt{\Delta_0(2\delta-\Delta_0)}$.
The expression for the pressure is
\begin{equation}
p_{\rm cond.}=-\frac{\bar\mu^2}{\pi^2}(2\delta-\Delta_0)^2.
\label{fifteenth}
\end{equation}
It corresponds to
the range $\Delta_0/2<\delta<\Delta_0$ for the displacement parameter.

3) The 2SC-LOFF state which is characterized
by nonzero values for $q$ and $\Delta$. Within
the LOFF window, $\delta_c^\prime<\delta<\delta_c$
with \cite{TI}
\begin{equation}
\delta_c^\prime\simeq 0.706\Delta_0\qquad \delta_c\simeq 0.754\Delta_0.
\label{sixteenth}
\end{equation}
The 2SC-LOFF state is both energetically favored and free from chromomagnetic 
instabilities \cite{GR2}.

The charge neutrality condition, 
\begin{equation}
\Big(\frac{\partial p}{\partial\mu_Q}\Big)_{\mu_B,\mu_8}
=\frac{1}{2}{\epsilon}
\Big(\frac{\partial p}{\partial\bar\mu}\Big)_{\delta,\delta^\prime}
-\Big(\frac{1}{3}+\frac{1}{2}\epsilon\Big)
\Big(\frac{\partial p}{\partial\delta}\Big)_{\bar\mu,\delta^\prime}=0
\label{neutral}
\end{equation}
gives rise to different values of the displacement
parameter $\delta$ for the normal and
different CSC phases for a given value of $\mu_B$. In the interesting
region where $\delta\sim\Delta_0$,
the impact of the different $\delta$'s on the pressure cannot 
be ignored within the weak
coupling approximation and the phase balance has to be
reconsidered. This task will be carried out
below. For the sake of analytical tractability, we shall
restrict our attention mainly to the
values of $\delta$ of the 2SC-LOFF phase
slightly below the threshold $\delta_c$. But we leave aside the 
possibility of a LOFF state with muti-plane waves.
The pressure from the LOFF condensate in this case reads
\begin{equation}
p_{\rm cond.}=\frac{4\bar\mu^2}{\pi^2}(\rho_c^2-1)(\delta-\delta_c)^2,
\label{eighteenth}
\end{equation}
with $\frac{q}{\delta}=\rho_c\simeq 1.20$ given by the equation
\begin{equation}
\frac{1}{\rho_c}\ln\frac{\rho_c+1}{\rho_c-1}=2.
\label{nineteenth}
\end{equation}
We shall continue to work 
with the grand canonical ensemble for fixed $\mu_B$ and we shall
retain the
approximation $\delta^\prime=0$. The difference between this approach
and the canonical ensemble with
fixed baryon number density as well as 
the correction that a nonzero $\delta^\prime$ induces, dictated
by the color neutrality are beyond the weak coupling approximation 
as we shall argue subsequently. 

In what follows, the pressure of a charge neutral state will be denoted 
by a capital letter $P$.

The pressure of the normal phase is approximated by (\ref{eighth}) 
at $\delta^\prime=0$, i.e.
\begin{equation}
p_n\simeq \frac{\bar\mu^2}{2\pi^2}
(\bar\mu^2+6\delta^2),
\label{twentith}
\end{equation}
where we have dropped higher order terms in $\delta$.
The charge neutrality condition (\ref{neutral}) yields
\begin{equation}
\delta=\frac{1}{2}\epsilon{\bar\mu}\equiv\delta_n
\label{Twfirst}
\end{equation}
and the pressure of the charge neutral normal phase becomes
\begin{equation}
P_n=\frac{\bar\mu^4}{2\pi^2}(1+\frac{3}{2}\epsilon^2).
\label{Twsecond}
\end{equation}

The pressure of the LOFF state is the sum of eq.(\ref{twentith}) and eq.
(\ref{eighteenth}) for $\delta$ close to
the upper edge of the LOFF window. Consequently we can write
\begin{equation}
p_{\rm LOFF}=p_n+\frac{4\bar\mu^2}{\pi^2}(\rho_c^2-1)
(\delta_c-\delta)^2.
\label{Twthird}
\end{equation}
The solution to the charge neutral condition (\ref{neutral}) is
\begin{equation}
\delta=\frac{3\delta_n+4(\rho_c^2-1)\delta_c}{4\rho_c^2-1}
\equiv\delta_{\rm LOFF}.
\label{Twforth}
\end{equation}
and we have
\begin{eqnarray}
P_{\rm LOFF} &=& P_n+\frac{9}{2}\Big(\frac{\partial^2p_n}{\partial\mu_Q^2}
\Big)_{\mu_B,\mu_8}\vert_{\mu_Q=-3\delta_n}(\delta_{\rm LOFF}-\delta_n)^2
+\frac{4\bar\mu^2}{\pi^2}(\rho_c^2-1)(\delta_c-\delta_{\rm LOFF})^2
\nonumber \\
&=& P_n +\frac{12\bar\mu^2}{\pi^2}\frac{\rho_c^2-1}{4\rho_c^2-1}
(\delta_c-\delta_n)^2.
\label{Twfifth}
\end{eqnarray}
Note that the contribution of the difference $\delta_{\rm LOFF}-
\delta_n$ to the pressure is of the same order as $p_{\rm cond.}$.
It follows from (\ref{sixteenth}) and (\ref{Twfirst}) that
\begin{equation}
\epsilon\bar\mu\simeq 1.508\Delta_0
\label{Twsixth}
\end{equation}
for $\delta_n\simeq\delta_c$.

The pressure of the 2SC phase is given by the sum of (\ref{twentith})
and eq.(\ref{fourteenth}), i.e.
\begin{equation}
p_{\rm 2SC}\simeq \frac{\bar\mu^2}{2\pi^2}(\bar\mu^2+2\delta^2+2\Delta_0^2).
\label{Twseventh}
\end{equation}
It follows from (\ref{neutral}) that
\begin{equation}
\delta=\frac{3}{2}\epsilon\bar\mu\equiv \delta_{\rm 2SC}.
\label{Tweighth}
\end{equation}
Comparing this state with the neutral 2SC-LOFF state by combining 
(\ref{Twsixth}) and (\ref{Tweighth})
we observe that $\delta_{\rm 2SC}=2.26\Delta_0$, which lies outside
the range of the 2SC solution. Stated differently, the 2SC state does 
not exist for the parametric relation (\ref{Twsixth}).

The pressure of the g2SC phase is given by the sum of (\ref{twentith}) and 
(\ref{fifteenth}). We find
\begin{equation}
p_{\rm g2SC}\simeq\frac{\bar\mu^2}{2\pi^2}(\bar\mu^2+8\Delta_0\delta
-2\delta^2-2\Delta_0^2].
\label{Twnineth}
\end{equation}
The charge neutrality (\ref{neutral}) yields
\begin{equation}
\delta=-\frac{3}{2}\epsilon\bar\mu+2\Delta_0\equiv\delta_{\rm g2SC}.
\label{Thirtith}
\end{equation}
Again comparing it with the neutral 2SC-LOFF at the same $\mu_B$, we find
that $\delta_{\rm g2SC}\simeq -0.26\Delta_0$ and the g2SC solution 
is similarly ruled out.

Therefore we have established that the neutral LOFF state is both
energetically 
favorable and chromomagnetically stable for a given baryonic chemical 
potential $\mu_B$ at $\mu_8=0$.

In a canonical ensemble approach
of the quark matter at a fixed baryon density, $n_B$,
the thermodynamic quantity which is minimized is the density of the
Helmholtz free energy
\begin{equation}
f = \mu_Bn_B-p
\label{Thfirst}
\end{equation}
subject to the condition that
\begin{equation}
n_B=\Big(\frac{\partial p}{\partial\mu_B}\Big)_{\mu_Q,\mu_8}=\frac{1}{3}
\Big(\frac{\partial p}{\partial\bar\mu}\Big)_{\delta,\delta^\prime},
\label{Thsecond}
\end{equation}
which generates slightly different $\mu_B$ for different phases. 
It follows from (\ref{nineth}) that 
the shift of $\mu_B$ in a super phase from that of the normal phase, i.e.
\begin{equation}
\delta\mu_B\equiv(\mu_B)_s-(\mu_B)_n\sim\Delta^2/(\bar\mu)_n
\label{Ththird}
\end{equation}
with 
\begin{equation}
(\bar\mu)_n\simeq \Big(\frac{3\pi^2n}{2}\Big)^{\frac{1}{3}}.
\end{equation} 
The term of order $\delta\mu_B$
is cancelled by the Legendre transformation 
on the RHS of (\ref{Thfirst}). The leading contribution of $\delta\mu_B$ 
to the Helmholtz free 
energy is of order $\mu_B^2\delta\mu_B^2\sim\Delta^4$ for all CSC states 
and therefore can be neglected. 

The color neutrality condition
\begin{equation}
\Big(\frac{\partial p}{\partial\delta^\prime}\Big)_{\bar\mu,\Delta,\delta}=0
\label{colorn}
\end{equation} 
yields $\delta^\prime=\mu_8=0$ in the normal phase
following the expression (\ref{eighth}). To asses its impact in 
the super phase, we have
to generalize our previous result, eq. (\ref{nineth}) for a nonzero 
$\delta^\prime$.
Since only red and green
quarks participate in the
pairing and a nonzero $\delta^\prime$ shifts their mean 
chemical potential without
offseting the Fermi momentum displacement, we have
\begin{equation}
p_{\rm cond.} = -\Gamma\Big(\bar\mu+\frac{\delta^\prime}{2\sqrt{3}},\Delta,
\delta,q\Big)
\end{equation}
A small value of 
\begin{equation}
\delta^\prime=\delta_s^\prime\sim\frac{\Delta^2}{\bar\mu}
\label{chem8}
\end{equation}
is induced in the super phase according to equ. (\ref{colorn})
and contributes a term of order $\Big(\frac{\partial^2 p_n}
{\partial\delta^{\prime 2}}\Big)\vert_{\delta^\prime
=0}\delta_s^{\prime 2}\sim\Delta^4$ to the pressure or the free energy of 
all CSC states, which is agian negligible.  

A sublety arises regarding the Meissner tensors of the 4-7th gluons.
In the previous paper \cite{GR2}, we derived
the small $\Delta/\delta$ expansion of
the Meissner mass tensors at $\delta^\prime=0$. We found that
\begin{equation}
(m_4^2)_{ij}=(m_5^2)_{ij}=(m_6^2)_{ij}=(m_7^2)_{ij}
=A(\delta_{ij}-\hat q_i\hat q_j)+B\hat q_i\hat q_j
\end{equation}
and 
\begin{equation}
(m_8^2)_{ij}=C(\delta_{ij}-\hat q_i\hat q_j)+D\hat q_i\hat q_j,
\end{equation}
where
\begin{equation}
A=\frac{g^2\bar\mu^2}{96\pi^2}\frac{1}
{(\rho_c^2-1)^2}\Big(\frac{\Delta}{\delta}\Big)^4 \ge 0,
\qquad B=\frac{g^2\bar\mu^2}{8\pi^2}\frac{1}{\rho_c^2-1}
\Big(\frac{\Delta}{\delta}\Big)^2 \ge 0
\label{meissner}
\end{equation}
and
\begin{equation}
C = 0, \qquad D=\frac{g^2\bar\mu^2}{6\pi^2}\Big(1+\frac{e^2}{3g^2}\Big)
\frac{1}{\rho_c^2-1}\Big(\frac{\Delta}{\delta}\Big)^2 \ge 0
\end{equation}
at equilibrium. The absence of a $\Delta^2$ term
in $A$ follows from the relation 
between the one loop diagram for the quadratic term of $A$ and that of 
the quadratic term of $C$ through an integration by parts when
$\delta^\prime=0$. The latter vanishes at equilibrium because of the 
second condition of eqs.(\ref{thirteenth}). This relation 
breaks down for a nonzero $\delta^\prime$ since the blue quarks enter
the one loop diagram of $A$ but not that of $C$. It can be shown explicitly 
that $A$ acquires a $\Delta^2$ term of order
$\bar\mu^2\Delta^2\delta^{\prime 2}/\delta^4$, which is suppressed relative 
to the expression of $A$ of eq.(\ref{meissner}) at weak coupling following 
the estimation (\ref{chem8}).
Therefore the property that the neutral 2SC-LOFF state is free from 
chromomagnetic instabilities remains intact under the color neutrality 
constraint.

In conclusion, we have found analytically a 2SC-LOFF state that is 
both color and electric neutral and free from chromomagnetic instabilities.
The controlling small parameter of the weak coupling approximation employed 
is the charge sum of the $u$ and 
$d$ flavors, which is of order $\Delta_0/\mu_B$ for the
above mentioned LOFF state. For realistic quark matter where the charge 
sum is $\frac{1}{3}$, we have 
\begin{equation}
\frac{\Delta_0}{\bar\mu}\simeq 0.22
\label{strength}
\end{equation}
in accordance with (\ref{Twsixth}). While the weak coupling approximation 
becomes marginal, it is still instructive to compare our result with 
the numerical analysis performed by Shovkovy and Huang \cite{SH}.
It appears that our case corresponds to the weaker NJL coupling
side of their results, where
the line of charge neutrality on the $\Delta-\delta$ plane excludes both 
the 2CS and g2SC states. We find that a
neutral LOFF state survives there. There are two possible future 
directions to follow. The first one is within 
the framework of the weak coupling approximation. One 
might attempt to extend the analysis to the lower value of the
displacement parameter $\delta$ and explore the phase balance of the 
2SC-LOFF and 
other exotic CSC states. In particular, because of different phases have
different values of
$\delta$, the LOFF window specified by (\ref{sixteenth}) will 
be modified. It would be very interesting if the window could become wider.
Alternatively, one may relax the week coupling approximation and 
perform a numerical analysis including the 2SC-LOFF state similar 
to that of Ref. \cite{SH}. 
In this manner more quantitative results can be obtained for realistic 
quark matter. 

After this work was completed, reference \cite{loff3}
appeared where a neutral LOFF state of three flavors is discussed.

\section*{Acknowledgments}
The work of I. G. and H. C. R. was supported in 
part by the US Department of Energy under grants
DE-FG02-91ER40651-TASKB while the work of D. H. was partly supported by
the NSFC under grant No. 10135030 and the Educational Committee of
China under grant No. 704035.

\end{document}